\documentclass[aps,pra,reprint,twocolumn,groupedaddress]{revtex4-2}
\usepackage{amsmath}
\usepackage{amssymb}
\usepackage{graphicx}
\usepackage{siunitx}
\usepackage{xcolor}
\newcommand{\Leg}[2]{\ensuremath{P_{#1,#2}}}

\begin{document}


\title{Shaped free electron vortices}


\author{D. Köhnke, T. Bayer, M. Wollenhaupt}
\affiliation{Carl von Ossietzky Universität Oldenburg, Institut für Physik, Carl-von-Ossietzky-Straße 9-11, D-26129 Oldenburg, Germany}


\date{\today}

\begin{abstract}
Since their first theoretical proposal \cite{NgokoDjiokap:2015:PRL:113004} and experimental demonstration \cite{Pengel:2017:PRL:053003}, free electron vortices (FEVs) have attracted considerable attention. 
Recently, a new class of unusually shaped FEVs, termed `reversible electron spirals', has been proposed theoretically \cite{Strandquist:2022:PRA:043110} for atomic single-photon ionization by two oppositely chirped and counterrotating circularly polarized (CRCP) attosecond pulses. Building on this concept, we present the first experimental demonstration of shaped FEVs created by atomic multiphoton ionization (MPI) using oppositely chirped CRCP femtosecond pulse pairs. In MPI by polarization-shaped pulses, several shaped FEVs of different rotational symmetry are superimposed resulting in the total photoelectron wave packet observed in the experiment. We employ velocity map imaging techniques to measure the photoelectron momentum distribution (PMD) and reconstruct the three-dimensional PMD by photoelectron tomography. Fourier analysis is used to decompose the measured PMD into the contributing FEVs with different rotational symmetries. The experimental results are supported by an analytical model and reproduced by numerical simulations, which provide insight into the role of resonant intermediate states. Our approach allows us to retrieve spectral information from the shaped laser field and the signatures of the MPI dynamics inscribed within the shaped FEVs.
\end{abstract}


\maketitle


\section{Introduction\label{Introduction}}
Strong-field ionization of atoms using polarization-shaped ultrashort laser pulses generally yields highly structured three-dimensional (3D) photoelectron wave packets \cite{Mancuso:2015:PRA:031402R,Eckart:2024:JPBAMOP:202001,Eckart:2016:PRL:133202,Hockett:2018b,Nauta:2020:OL:2156,Hasovic:2016:OE:6413,Wollenhaupt:2013:CPC:1341}. 
In a multiphoton picture, these wave packets arise from the interference of several angular momentum partial wave packets in the continuum, created via the different multiphoton ionization (MPI) pathways resulting from the dipole selection rules \cite{Wollenhaupt:2009:APB:245,Hockett:2014:PRL:223001,Kerbstadt:2019:NC:658}. Free electron vortices (FEVs) are a prime example of such 3D structured photoelectron wave packets \cite{NgokoDjiokap:2015:PRL:113004,Pengel:2017:PRL:053003,Kerbstadt:2019:NC:658}. FEVs have recently attracted considerable attention as prototype models for studying the interaction of polarization-tailored light fields with atoms \cite{Pengel:2017:PRA:043426,NgokoDjiokap:2016:PRA:013408,Li:2017:COL:120202,Kong:2018:JOSAB:2163,Jia:2019:CPL:119,Zhen:2020:CPL:136885,Armstrong:2019:PRA:063416,Eickhoff:2021:FP:444,Koehnke:2023:NJP:123025} or molecules \cite{Yuan:2017:JPB:124004,Djiokap:2018:PRA:063407,Guo:2021:LP:065301,Bayer:2022:FC:899461,Wang:2023:JOSAB:1749}.
Moreover, FEVs were shown to be highly sensitive probes of light-matter interactions, providing detailed interferometric information on the underlying excitation and ionization dynamics \cite{Yuan:2016:PRA:053425,Djiokap:2017:JO:124003,Djiokap:2018:PRA:063407,Li:2018:OE:878,Ke:2019:OE:32193,Xiao:2019:PRL:053201,Qin:2020:FP:32502,Djiokap:2021:PRA:053110,Bayer:2022:FC:899461,Wang:2023:JOSAB:1749}, including the influence of intermediate resonances and relative quantum phases of different photoionization continua \cite{Koehnke:2022:JPBAMOP:184003}. \\
Initially, FEVs were created by the perturbative atomic photoionization with a sequence of two transform-limited (TL) counterrotating circularly polarized (CRCP) ultrashort laser pulses separated in time by a delay $\tau$. Ramsey-type interference of the two sequentially released angular momentum partial wave packets with different magnetic quantum numbers $m$ gives rise to a spiral-shaped photoelectron momentum distribution (PMD), characterized by a distinct rotational symmetry in the laser polarization plane (azimuthal plane; angle $\phi$). The rotational symmetry, i.e. number of spiral arms, is given by the difference $\Delta m$ between the magnetic quantum numbers of the interfering partial wave packets.  In the case of TL pulses, the FEVs are Archimedean spirals: The spiral arms, defined as the lobes of the interference pattern, are linear functions of the photoelectron kinetic energy $\varepsilon$. The slope $\tau/\hbar$ is determined by the time-delay between the two partial wave packets \cite{Pengel:2017:PRA:043426}. More generally, the shape $\phi(\varepsilon)$ of the spiral arms maps the spectral phase of the multiphoton spectrum of the ionizing laser pulses. \\
Since their first proposal \cite{NgokoDjiokap:2015:PRL:113004} and experimental demonstration \cite{Pengel:2017:PRL:053003}, numerous studies on different types of FEVs have been reported. For example, FEVs with odd rotational symmetry have been created by MPI of sodium atoms using TL bichromatic pulses \cite{Kerbstadt:2019:NC:658}. Non-perturbative FEVs have been demonstrated by atomic resonance-enhanced multiphoton ionization (REMPI) of potassium atoms using TL $\pi$-pulses \cite{Pengel:2017:PRL:053003,Pengel:2017:PRA:043426}. FEVs created by double ionization of helium atoms with TL pulses have been investigated theoretically in \cite{NgokoDjiokap:2017:PRA:013405}, and the creation of molecular FEVs by MPI with TL single-color and bichromatic pulses has been studied in \cite{Yuan:2016:PRA:053425,Djiokap:2018:PRA:063407,Guo:2021:LP:065301,Bayer:2022:FC:899461}. \\ 
In this work, we go beyond previous studies by investigating the creation of FEVs by atomic photoionization with \textit{shaped} CRCP femtosecond (fs) pulses.   
This new class of photoelectron wave packets is referred to as \textit{shaped} electron vortices (SEVs). In coherent control, linearly chirped pulses are the prime examples of pulse tailoring. Therefore, we use oppositely chirped CRCP pulses generated by quadratic spectral phase modulation of the two CRCP pulse components for the initial experimental demonstration of SEVs.
Very recently, the creation of $c_2$-symmetric SEVs by atomic single-photon ionization with oppositely chirped CRCP attosecond pulses was proposed theoretically by Strandquist \textit{et al.} \cite{Strandquist:2022:PRA:043110}. Here we extend this concept and investigate the creation of SEVs by atomic MPI with oppositely chirped CRCP fs pulses. The photoelectron wave packets created by MPI are generally more complex than those produced by single-photon ionization due to the rich MPI dynamics. In MPI, multiple pathways give rise to an intricate interference pattern of several angular momentum partial wave packets in the continuum, each of which has a different energy spectrum. In the perturbative regime, the spectral amplitudes and phases of the created angular momentum partial wave packets are determined by the multiphoton spectra associated with the corresponding MPI pathways \cite{Eickhoff:2021:JPB:164002}, which we refer to as the pathway-resolved multiphoton spectra in the following. Thus, polarization-shaped laser pulses imprint different energy-dependent, non-linear quantum phases on each partial wave packet. The pairwise interference of partial wave packets with different $m$ then gives rise to multiple SEVs with different rotational symmetries and non-Archimedean shapes. Eventually, the superposition of all the SEVs results in the total photoelectron wave packet, whose probability density is measured in the experiment. \\
In this paper, we present the first experimental demonstration of SEVs created by the (1+2) REMPI of potassium atoms using oppositely chirped CRCP fs-pulses.
Specifically, we observe a PMD consisting of the superposition of SEVs with $c_6$-,  $c_4$- and $c_2$-symmetry. Due to the chirp, the spiral arms have a quadratic shape $\phi(\varepsilon)\propto\varepsilon^2$,
implying that the direction of rotation of the SEVs reverses at a well-defined photoelectron kinetic energy, as suggested by the term `reversible electron spirals' coined in \cite{Strandquist:2022:PRA:043110}. Our observations are qualitatively well reproduced by a simple analytical model based on the perturbative MPI of a single-electron atom, and are in excellent agreement with numerical simulations taking into account the essential intermediate states. Experimentally, we combine white-light polarization pulse shaping \cite{Brixner:2001:OL:557,Wollenhaupt:2009:APB:245,Kerbstadt:2017:OE:12518} with velocity map imaging (VMI) \cite{Eppink:1997:RSI:3477,Whitaker:2003:249} for the differential measurement of the PMD. The 3D PMD is reconstructed using the photoelectron tomography technique \cite{Wollenhaupt:2009:APB:647,Smeenk:2009:JPB:185402,Wollenhaupt:2013:CPC:1341}.
To disentangle the different SEVs superimposed in the total PMD according to their azimuthal symmetry, we developed a Fourier analysis method which is specifically adapted to the symmetry properties of the angular momentum wave functions. Analyzing our experimental data allows us to determine the quantum phase of the SEVs. The result confirms that the SEV-phases are essentially determined by the optical phases of the pathway-resolved multiphoton spectra. We also find signatures of resonant intermediate states in the MPI, which are imprinted as phase-jumps in the SEV-phases. 

\section{Physical system and model\label{PhysicalSystem}}
\begin{figure}[ht]
	\includegraphics[width=\linewidth]{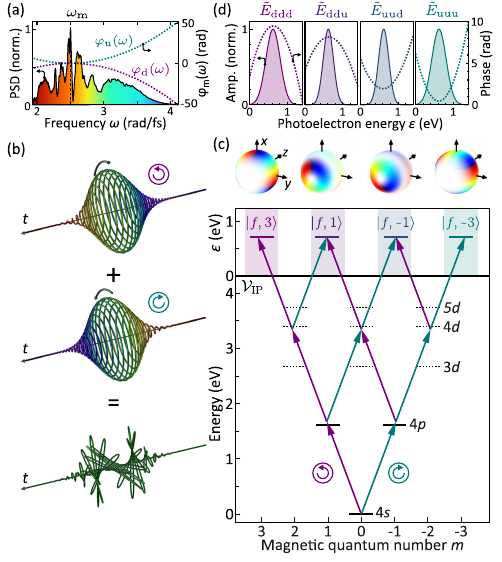}
	\caption{Physical system. The white-light supercontinuum shown in (a) is decomposed into two CRCP pulse components which are chirped by quadratic spectral phase-modulation of opposite sign. Superposition of the two CRCP components yields a pulse with a twisted polarization profile, as shown in (b). The color-coding of the curves indicates the corresponding instantaneous frequencies. (c) Excitation scheme for the MPI of potassium atoms by the tailored pulse. Different combinations of LCP and RCP photons open up MPI pathways to four different angular momentum channels in the ionization continuum. The angular parts of the created angular momentum partial wave packets are indicated on top. The corresponding kinetic energy distributions given by the pathway-resolved multiphoton spectra are shown in (d).\label{fig1}}
\end{figure}
In this section, we develop an analytical model to describe the SEVs created by the perturbative MPI of potassium atoms with oppositely chirped CRCP fs-laser pulses. In particular, the model describes the relation between the SEVs and the 3D PMD measured in the experiment, and  provides the basis for the Fourier analysis method described in Sec.~\ref{subsec:Fourier} and applied in Sec.~\ref{Results} to decompose the measured PMD into the SEVs. To this end, we describe the spectrum of oppositely chirped CRCP femtosecond fs-pulses in Sec.~\ref{PhysicalSystemPulse} for the perturbative MPI of potassium atoms described in Sec.~\ref{PhysicalSystemMPI}.

\subsection{Shaped laser pulse \label{PhysicalSystemPulse}}
We start by considering the polarization-shaped laser pulse. 
The pulse is created by superimposing two temporally overlapping polarization components of opposite handedness, i.e., a left-handed (LCP) and a right-handed circularly polarized (RCP) component. Each component is linearly chirped with the same absolute chirp parameter but opposite sign. 
The chirp is introduced by spectral phase-modulation using quadratic phase functions of the form
\begin{equation}\label{eq:chirp} \varphi_{\mathrm{u}/\mathrm{d}}(\omega)=\frac{\phi^{(2)}}{2}\,(\omega-\omega_\mathrm{m})^2,
\end{equation}
centered around the frequency $\omega_\mathrm{m}$ chosen close to the central frequency $\omega_0$ of the pulse spectrum $\tilde{E}(\omega)$. A positive (negative) chirp parameter $\phi^{(2)}$ in Eq.~\eqref{eq:chirp} corresponds to an up-chirped (down-chirped) component.
Since the chirp parameters of the RCP and the LCP component are equal in magnitude, it is sufficient to specify the chirp parameter of the RCP component.
Then, for $\phi^{(2)}>0$, the vectorial electric field of the oppositely chirped CRCP pulse, described in the frequency domain and in the spherical basis, reads
\begin{align}
	\boldsymbol{\tilde{\mathcal{E}}}^+(\omega)&=\tilde{\mathcal{E}}(\omega)\left[e^{-\mathrm{i}\varphi_{\mathrm{u}}(\omega)}\mathbf{e}_{-1}+e^{-\mathrm{i}\varphi_{\mathrm{d}}(\omega)}\mathbf{e}_{+1}\right]\notag\\
	&=\tilde{\mathcal{E}}_\mathrm{u}(\omega)\mathbf{e}_{-1}+\tilde{\mathcal{E}}_\mathrm{d}(\omega)\mathbf{e}_{+1}.
	\label{eq:Spectrum}
\end{align} 
Herein, $\tilde{\mathcal{E}}(\omega)$ is the real-valued spectral envelope common to both polarization components, $\mathbf{e}_{\pm1}$ are the spherical unit vectors describing LCP and RCP light, respectively, and $\tilde{\mathcal{E}}_\mathrm{u/d}(\omega)$ are the corresponding complex-valued spectral envelopes. For $\phi^{(2)}<0$, the assignment between chirp and polarization is interchanged, i.e., the RCP component is down-chirped and the LCP component is up-chirped.
The measured fundamental spectrum $|\tilde{E}^+(\omega)|^2=|\tilde{\mathcal{E}}(\omega-\omega_0)|^2$ of the two pulse components is shown in Fig.~\ref{fig1}(a), together with the applied spectral phase functions $\varphi_\mathrm{u/d}(\omega)$.
The resulting pulse in time domain is illustrated in Fig.~\ref{fig1}(b) superimposing two Gaussian-shaped oppositely chirped CRCP pulse components. The pulses exhibit a twisted polarization profile. At any time $t$, the electric field of the pulse is approximately linearly polarized. However, its direction of polarization rotates continuously with a time-varying angular velocity. In the rising edge of the pulse, the rotation decelerates up to $t=0$, where it comes to a halt, reverses its direction and accelerates again in the falling edge of the pulse. The instantaneous frequency of the polarization-shaped pulse, given by the mean value of the instantaneous frequencies of the two oppositely chirped pulse components \cite{Wollenhaupt:2015:63}, is constant with $\omega_{\mathrm{inst}}(t)=\omega_0$.

\subsection{MPI of potassium atoms\label{PhysicalSystemMPI}}
\begin{figure}[t]
	\includegraphics[width=\linewidth]{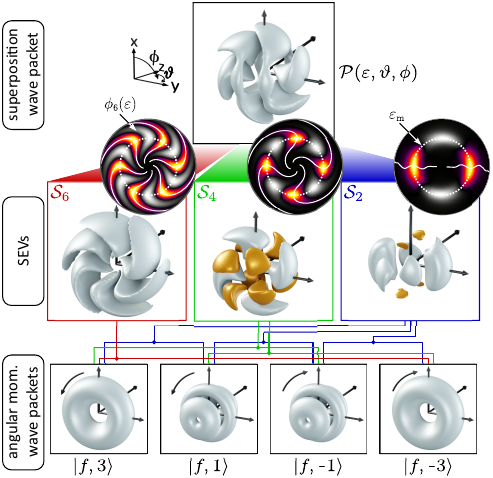}
	\caption{Schematic (de-)composition of the PMD observed in the experiment (upper row). The PMD is composed of three SEVs with $c_6$-, $c_4$- and $c_2$-rotational symmetry (central row). The SEVs are interferograms resulting from the pairwise interference of different angular momentum partial wave packets (lower row) created along the different MPI pathways. 
	The interference pattern of the SEVs in the polarization plane ($x$-$y$-plane), shown in the circular insets, reflect the energy-dependent quantum phases of the SEVs in the shape of the vortex arms (solid white lines).
	Due to their rotational symmetry, the SEVs can be retrieved by Fourier analysis of the PMD in azimuthal direction.
	\label{fig2}}
\end{figure}

Next we consider the interaction of the polarization-shaped pulse in Eq.~\eqref{eq:Spectrum} with potassium atoms. The corresponding excitation scheme is shown in Fig.~\ref{fig1}(c). According to the selection rules for optical dipole transitions, the LCP component drives $\sigma^+$-transitions (violet arrows), while the RCP component drives $\sigma^-$-transitions (turquoise arrows). Hence, starting from the $|4s,m=0\rangle$ ground state, the absorption of different combinations of LCP and RCP photons opens up multiple MPI pathways to different angular momentum target states $|\ell,m\rangle$ in the continuum. For simplicity, we apply the propensity rule $\Delta \ell=+1$ and consider an $f$-type continuum. The corresponding photoelectron partial wave packets are referred to as angular momentum wave packets. Their angular distributions are described by the spherical harmonics $Y_{\ell,m}(\vartheta,\phi)$, as illustrated on top of Fig.~\ref{fig1}(c). In the third-order time-dependent perturbation description, their radial parts, i.e., their energy spectra, are determined by superpositions of the pathway-resolved three-photon spectra $\tilde{\mathcal{E}}_{\alpha\beta\gamma}(\varepsilon)=\mathcal{F}[\mathcal{E}_\alpha \mathcal{E}_\beta\mathcal{E}_\gamma](\varepsilon)$, with $\alpha,\beta,\gamma\in\{\mathrm{u},\mathrm{d}\}$, associated with the MPI pathways leading to the corresponding target state. In general, i.e. for arbitrarily shaped pulses, these radial parts are strongly modulated in both amplitude and phase. Here, we consider oppositely chirped pulses for the demonstration of SEVs in MPI.
Assuming a Gaussian-shaped, linearly chirped fundamental spectrum with a bandwidth $\Delta\omega$, the pathway-resolved multiphoton spectra are likewise Gaussian-shaped and linearly chirped, as shown in Fig.~\ref{fig1}(d). The effective bandwidth $\Delta\omega_\mathrm{\alpha\beta\gamma}$ and the effective chirp parameter $\phi_{\mathrm{\alpha\beta\gamma}}$ of the multiphoton spectra depend on the associated MPI pathways and are therefore different for each target state. Because all pathways leading to a particular target state are composed of the same combination of photons -- in different permutations -- all of these pathways have the same multiphoton spectrum. The analytical results for the effective bandwidth $\Delta\omega_\mathrm{\alpha\beta\gamma}$ and the effective chirp parameter $\phi_{\mathrm{\alpha\beta\gamma}}$ obtained for a Gaussian-shaped pulse are given in Tab.~\ref{tab1} and derived in Appendix~\ref{App:Pathway_resolved}.
\begin{table}
	\renewcommand{\arraystretch}{1.5}
	\setlength{\tabcolsep}{0.5cm}
	\caption{Effective multiphoton bandwidths $\Delta\omega_\mathrm{\alpha\beta\gamma}$ and chirp parameters $\phi_{\mathrm{\alpha\beta\gamma}}$ of the different pathway-resolved multiphoton spectra produced by a Gaussian-shaped fundamental of bandwidth $\Delta\omega$ and linear chirp parameter $\phi^{(2)}$. The dimensionless parameter 
	$\eta= \left[ \ln(16) \, \frac{\phi^{(2)}}{\Delta t^2} \right]^2 =\left[\frac{\Delta\omega^2\phi^{(2)}}{\ln(16)}\right]^2$ is introduced for compact notation. \label{tab1}}	
	\begin{tabular}{ccc}
		$\alpha\beta\gamma$&$\Delta\omega_\mathrm{\alpha\beta\gamma}$&$\phi_{\mathrm{\alpha\beta\gamma}}$\\
		\hline\hline
		uuu&$\sqrt{3}\,\Delta\omega$&$\frac{1}{3}\phi^{(2)}$\\
		uud&$\sqrt{\frac{1}{3}\frac{\eta+9}{\eta+1}}\Delta\omega$&$\frac{\eta+1}{\eta+9}\phi^{(2)}$\\
		udd&$\sqrt{\frac{1}{3}\frac{\eta+9}{\eta+1}}\Delta\omega$&$\quad-\frac{\eta+1}{\eta+9}\phi^{(2)}\quad$\\
		ddd&$\sqrt{3}\,\Delta\omega$&$-\frac{1}{3}\phi^{(2)}$\\
		\hline\hline
	\end{tabular}
\end{table}
The two `pure' three-photon spectra $\tilde{\mathcal{E}}_{\mathrm{uuu}}(\omega)$ and $\tilde{\mathcal{E}}_{\mathrm{ddd}}(\omega)$, which result from the absorption of either three RCP photons from the up-chirped or three LCP photons from the down-chirped pulse component, have the same effective bandwidth of $\sqrt{3}\Delta\omega$ regardless of the chirp parameter. In contrast, the effective bandwidth of the two mixed three-photon spectra $\tilde{\mathcal{E}}_{\mathrm{uud}}(\omega)$ and $\tilde{\mathcal{E}}_{\mathrm{udd}}(\omega)$, which result from the absorption of photons from both polarization components, is always smaller for a given $\phi^{(2)}\neq0$ and tends asymptotically towards a lower limit of $\frac{1}{\sqrt{3}}\Delta\omega $ for very large values of $|\phi^{(2)}|$. The effective chirp parameters of the `pure' multiphoton spectra $\tilde{\mathcal{E}}_{\mathrm{uuu}}(\omega)$ and $\tilde{\mathcal{E}}_{\mathrm{ddd}}(\omega)$ are $\pm\frac{1}{3}\phi^{(2)}$, respectively. For the mixed multiphoton spectra $\tilde{\mathcal{E}}_{\mathrm{uud}}(\omega)$ and $\tilde{\mathcal{E}}_{\mathrm{udd}}(\omega)$ the effective chirp parameters tend towards $\pm\frac{1}{9}\phi^{(2)}$ for very small $|\phi^{(2)}|$ and towards $\pm\phi^{(2)}$ for very large $|\phi^{(2)}|$, respectively.\\
The densities of the angular momentum wave packets created by a Gaussian-shaped \SI{6}{fs} pulse, are illustrated in the bottom row of Fig.~\ref{fig2} sorted by $m$. The pairwise interference of the four angular momentum wave packets with different $m$ gives rise to three different SEVs with $c_6$, $c_4$ and $c_2$ rotational symmetry in azimuthal direction (angle $\phi$). Since the $c_6$-symmetric SEV arises from two ionization pathways, it is physically analogous to the $c_2$-symmetric reversible electron spiral from single-photon ionization described in \cite{Strandquist:2022:PRA:043110}. These SEVs, obtained for a chirp parameter of $\phi^{(2)}=\SI{40}{fs^2}$, are shown in the middle row of Fig.~\ref{fig2}. Their rotational symmetry is determined by the difference of the involved magnetic quantum numbers. The shapes of the spiral arms follow quadratic functions $\phi_j(\varepsilon)\propto\varepsilon^2$. The corresponding curves are represented by solid lines in the 2D sections through the SEVs in the $x$-$y$-plane at $z=0$, shown in the insets. Note that the section of the $c_2$-symmetric SEV has a more complex structure. In this case, the pairwise interferences lead to two different spectral phases being mapped into the SEV. These phases mix with different weights in the polar direction. Below, in Sec.~\ref{Results}, we show how to disentangle both contributions by analyzing the polar dependence of the SEV in addition. The energy of reversal $\varepsilon_\mathrm{m}$ of the spiral arms, indicated as dotted circle in the circular insets to Fig.~\ref{fig2}, is determined by the modulation frequency $\omega_\mathrm{m}$ as
\begin{equation}\label{eq:eor} \varepsilon_\mathrm{m}=3\hbar\omega_\mathrm{m}-\mathcal{V}_\mathrm{IP}, 
\end{equation}
where $\mathcal{V}_\mathrm{IP}$ denotes the ionization potential (IP) of the atom. Finally, the coherent superposition of all SEVs results in the superposition wave packet, whose density $\mathcal{P}(\varepsilon,\vartheta,\phi)$ is shown in the top row of Fig.~\ref{fig2} and measured as the 3D PMD in the experiment. In the perturbative regime, we obtain 
\begin{align}\label{eq:ShapedVortex}
	\mathcal{P}(\varepsilon,\vartheta,\phi)=&\;\left|\tilde{\mathcal{E}}_\mathrm{uuu}Y_{3,3}+\tilde{\mathcal{E}}_\mathrm{uud}Y_{3,1}\right.\notag\\		&\left.\;+\,\tilde{\mathcal{E}}_\mathrm{udd}Y_{3,-1}+\tilde{\mathcal{E}}_\mathrm{ddd}Y_{3,-3}\right|^2\notag\\
		=&\;\mathcal{S}_0+\mathcal{S}_2+\mathcal{S}_4+\mathcal{S}_6.
\end{align}
The first term $\mathcal{S}_0=\mathcal{S}_0(\varepsilon,\vartheta)$ describes an azimuthally isotropic background, while the terms $\mathcal{S}_j=\mathcal{S}_j(\varepsilon,\vartheta,\phi)$, with $j\in\{2,4,6\}$, describe the SEVs with $c_j$-rotational symmetry in the $\phi$-direction. Since the background $\mathcal{S}_0$ is considered as a separate term, the $\mathcal{S}_j$ are not strictly positive, i.e. their values are centered around zero. Therefore, the SEVs are not densities, but rather interferograms.
Explicitly, the SEVs are of the following form, where the arguments $\varepsilon,\vartheta,\phi$ are omitted for brevity:
\begin{align}
	\mathcal{S}_6=&\;-\mathcal{A}_6\Leg{3}{3}^2\cos\left(6\phi+\varphi_6\right),\label{eq:SEV6}\\
	\mathcal{S}_4=&\;-2\mathcal{A}_4\Leg{3}{3}\Leg{3}{1}\cos\left(4\phi+\varphi_4\right),\label{eq:SEV4}\\
	\mathcal{S}_2=&\;2\mathcal{A}_{21}\Leg{3}{3}\Leg{3}{1}\cos\left(2\phi+\varphi_{21}\right)\notag\\
	&-\mathcal{A}_{22}\Leg{3}{1}^2\cos\left(2\phi+\varphi_{22}\right),\label{eq:SEV2}
\end{align}
where $\Leg{\ell}{m}=\Leg{\ell}{m}[\cos(\vartheta)]$ are the normalized associated Legendre polynomials.  
The amplitudes $\mathcal{A}_j=\mathcal{A}_j(\varepsilon)$ and spectral phase functions $\varphi_j=\varphi_j(\varepsilon)$ of the SEVs are determined by the products of the pathway-resolved multiphoton spectra as
\begin{align}
	\mathcal{A}_6(\varepsilon)&=|\tilde{\mathcal{E}}_\mathrm{uuu}\tilde{\mathcal{E}}^*_\mathrm{ddd}|,&\varphi_6(\varepsilon)&=\arg\{\tilde{\mathcal{E}}_\mathrm{uuu}\tilde{\mathcal{E}}^*_\mathrm{ddd}\} \label{eq:phi6}\\
	\mathcal{A}_4(\varepsilon)&=|\tilde{\mathcal{E}}_\mathrm{uuu}\tilde{\mathcal{E}}^*_\mathrm{udd}|,&\varphi_4(\varepsilon)&=\arg\{\tilde{\mathcal{E}}_\mathrm{uuu}\tilde{\mathcal{E}}^*_\mathrm{udd}\}\label{eq:phi4}\\
	\mathcal{A}_{21}(\varepsilon)&=|\tilde{\mathcal{E}}_\mathrm{uuu}\tilde{\mathcal{E}}^*_\mathrm{uud}|,&\varphi_{21}(\varepsilon)&=\arg\{\tilde{\mathcal{E}}_\mathrm{uuu}\tilde{\mathcal{E}}^*_\mathrm{uud}\}\label{eq:phi21}\\
	\mathcal{A}_{22}(\varepsilon)&=|\tilde{\mathcal{E}}_\mathrm{uud}\tilde{\mathcal{E}}^*_\mathrm{udd}|,&\varphi_{22}(\varepsilon)&=\arg\{\tilde{\mathcal{E}}_\mathrm{uud}\tilde{\mathcal{E}}^*_\mathrm{udd}\}.\label{eq:phi22}
\end{align}
In the case of a Gaussian-shaped, linearly chirped fundamental spectrum discussed above, the amplitudes are products of Gaussian-shaped functions with different bandwidths and the spectral phases are also quadratic functions, i.e. the pathway-resolved multiphoton spectra are chirped pulses with the  effective bandwidth $\Delta\omega_\mathrm{\alpha\beta\gamma}$ and the effective chirp parameter $\phi_{\mathrm{\alpha\beta\gamma}}$ given in Tab.~\ref{tab1}. 
Hence, the spectral phase functions are given by
\begin{align}
	\varphi_6(\varepsilon) &= \frac{2}{3}\phi^{(2)}\frac{(\varepsilon-\varepsilon_\mathrm{m})^2}{\hbar^2} \label{eq:phi6_chirp} \\
	\varphi_4(\varepsilon) &= \left(\frac{1}{3}+\frac{\eta+1}{\eta+9}\right)\phi^{(2)}\frac{(\varepsilon-\varepsilon_\mathrm{m})^2}{\hbar^2}\label{eq:phi4_chirp}\\
	\varphi_{21}(\varepsilon) &= \left(\frac{1}{3}-\frac{\eta+1}{\eta+9}\right)\phi^{(2)}\frac{(\varepsilon-\varepsilon_\mathrm{m})^2}{\hbar^2}\label{eq:phi21_chirp}\\
	\varphi_{22}(\varepsilon) &= 2\frac{\eta+1}{\eta+9}\phi^{(2)}\frac{(\varepsilon-\varepsilon_\mathrm{m})^2}{\hbar^2}\label{eq:phi22_chirp}
\end{align}

\subsection{3D Fourier analysis\label{subsec:Fourier}}
In the multiphoton picture, the PMD is a superposition of the angular momentum wave packets created along the different MPI pathways. Since the azimuthal part of the angular momentum wave packets reads $e^{\mathrm{i}m\phi}$, Eq.~\eqref{eq:ShapedVortex} describes the density $\mathcal{P}(\varepsilon,\vartheta,\phi)$ of the superposition wave packet as a Fourier synthesis of the azimuthal standing waves $\cos[j\, \phi + \Phi_j(\varepsilon,\vartheta)]$, which are associated with the $c_j$-symmetric SEVs with amplitudes $\mathcal{C}_j(\varepsilon,\vartheta)$ and phases $\Phi_j(\varepsilon,\vartheta)$ in Eqns.~\eqref{eq:SEV6}-\eqref{eq:SEV2}:  
\begin{equation}
\mathcal{P}(\varepsilon,\vartheta,\phi) = \sum_{j=0,2,4,6} \mathcal{C}_j(\varepsilon,\vartheta) \cos[j\,\phi + \Phi_j(\varepsilon,\vartheta)].
\label{eq:Fouriersynthesis}
\end{equation}
It is straightforward to extract the individual SEVs from the measured PMD by performing a Fourier analysis of the PMD in the azimuthal direction.
Applying the Fourier analysis for each energy $\varepsilon$ and polar angle $\vartheta$, we obtain the 2D amplitudes $\mathcal{C}_j(\varepsilon,\vartheta)$ and the spectral phase functions $\Phi_j(\varepsilon,\vartheta)$ for each $c_j$-symmetry component. The corresponding 3D SEVs $\mathcal{S}_j(\varepsilon,\vartheta,\phi)$ are then reconstructed by inverse Fourier transform of the $j$-th symmetry component.  Therefore, we refer to the method as a 3D Fourier analysis. A change of the sign of the $\vartheta$-dependent amplitudes described by the associated Legendre polynomials in Eqns.~\eqref{eq:SEV6}-\eqref{eq:SEV2} results in a $\pi$ phase-jump in the polar direction of the retrieved phase functions $\Phi_j(\varepsilon,\vartheta)$. These phase-jumps, highlighted in the middle row of Fig.~\ref{fig2} by the color-coding of the SEV isosurfaces, reveal details of the angular structure of the target states involved. For example, the change of sign of the $|f,\pm1\rangle$-type angular momentum wave packets leads to $\pi$ phase jumps in polar direction. By describing the radial part of the angular momentum wave packets by the multiphoton spectra, we neglect the MPI dynamics due to the time-evolution of the electron wave function in the resonant $4p$-state and the near-resonant higher-lying $d$- and $s$-states \cite{Eickhoff:2022:PRA:053113}. As a consequence, deviations of $\Phi_j(\varepsilon,\vartheta)$ in the retrieved data from the multiphoton spectral phases described by Eqns.~\eqref{eq:phi6}-\eqref{eq:phi22} reveal signatures of the transient electronic excitation dynamics in these intermediate states. 
Our numerical simulation of the (1+2) REMPI scenario includes all essential states to fully capture the MPI dynamics. 

\section{Experimental setup\label{Setup}}
In our experiment, we combine fs white-light polarization pulse shaping \cite{Brixner:2001:OL:557,Weiner:2011:OC:3669,Kerbstadt:2017:JMO:1010} with VMI-based \cite{Eppink:1997:RSI:3477} photoelectron tomography \cite{Wollenhaupt:2013:CPC:1341,Smeenk:2009:JPB:185402}. The primary light source is a chirped pulse amplification system (\textsc{FEMTOLASERS RAINBOW 500}, \textsc{CEP4 Module}, \textsc{FEMTOPOWER HR CEP}) with a central wave length of \SI{790}{\nano\meter}, a repetition rate of \SI{3}{kHz} and a pulse energy of \SI{1.0}{\milli\joule}. The infrared laser pulses are used to seed a neon-filled hollow-core fiber at an absolute pressure of \SI{2}{\bar} for the generation of a white-light supercontinuum (WLS), as shown in Fig.~\ref{fig1}(a). Using a $4f$ polarization pulse shaper based on a dual-layer liquid crystal spatial light modulator (LC-SLM, \textsc{Jenoptik SLM-S640d}), the spectral phases of two orthogonal linearly polarized WLS components are modulated by quadratic spectral phase functions (cf. Sec.~\ref{PhysicalSystem}). A superachromatic $\lambda/4$ waveplate is used to convert the orthogonal linearly polarized components into the CRCP components, resulting in the polarization-shaped output pulse. In addition, a $\lambda/2$ waveplate is used to rotate the output pulse for the tomographic reconstruction of the 3D PMD \cite{Wollenhaupt:2009:APB:647}. \\
The polarization-shaped pulses are focused by a spherical mirror ($f=\SI{250}{\milli\meter}$) into the interaction region of a VMI spectrometer filled with potassium vapor from a dispenser source (SAES Getters). The photoelectron wave packets created by atomic MPI are projected onto a micro-channel plate which amplifies the photoelectron signal for visualization on a phosphor screen. The image of the screen is recorded by a charge-coupled device camera (CCD; Lumenera LW165M) using an exposure time of \SI{400}{\milli\second}. To tomographically reconstruct the PMD, the $\lambda/2$ waveplate is rotated in increments of $\SI{2}{\degree}$ from $\SI{0}{\degree}$ to \SI{90}{\degree}, effectively rotating the PMD by \SI{180}{\degree}. Each projection is measured by averaging over 200 CCD images for each PMD orientation. The 3D PMD is obtained from the recorded 45 projections using the Fourier slice algorithm \cite{Kak:1988:1}.
\section{Results\label{Results}}
\begin{figure}[t]
	\includegraphics[width=\linewidth]{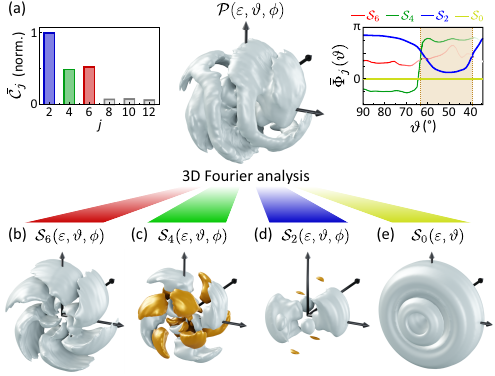}
	\caption{Analysis of the superpositinon wave packet measured for a chirp parameter of $\phi^{(2)}=\SI{40}{fs^2}$. (a) Measured and tomographically reconstructed density of the superposition wave packet. The upper left inset shows the amplitude distribution obtained after Fourier decomposition of the PMD into its azimuthal symmetry components. 
	(b)-(d) 3D representations of the predominant components $c_6$, $c_4$ and $c_2$. The interferograms exhibit intricate spiral shapes corresponding to the SEVs $\mathcal{S}_6$, $\mathcal{S}_4$ and $\mathcal{S}_2$, respectively. The SEVs are color-coded according to the phase-jumps observed in polar ($\vartheta$) direction, as shown in the upper right inset in (a). (e) Azimuthally isotropic background $\mathcal{S}_0$, corresponding to the DC component of the PMD.\label{fig3}}
\end{figure}
To demonstrate SEVs by MPI of potassium atoms in the experiment, we applied quadratic spectral phase-modulation functions of opposite sign to the orthogonal WLS polarization components.
Starting with a fixed chirp parameter, we examine the PMD with respect to the superimposed SEVs. Next, we manipulate the SEVs by varying the sign and magnitude of the chirp parameter. Finally, we analyze the shape of the SEVs in detail, retrieve their energy-dependent phases and compare the experimental findings to the analytical model and our simulations. \\
The modulation frequency $\omega_\mathrm{m}=\SI{2.50}{\radian\per\femto\second}$ of the two chirps, defined in Eq.~\eqref{eq:chirp}, was chosen so that the energy of reversal of the spiral arms, described by Eq.~\eqref{eq:eor}, was located near the center of the photoelectron energy spectrum at 
$\varepsilon_m\approx\SI{0.6}{\electronvolt}$. The TL pulse duration of the WLS pulses is $\Delta t=\SI{6}{\femto\second}$. The intensity of the laser pulses is estimated to be about \SI{4e12}{\watt\per\centi\meter^2}, ensuring perturbative interaction conditions \cite{Karule:1990:265}. The measured 2D projections of the PMD were symmetrized along the $y$- and the $z$-direction, since PMDs created by single-color MPI are expected to be left-right and forward-backward symmetric with respect to the laser propagation direction \cite{Eickhoff:2021:FP:444}. The redundancy of the experimental data was utilized to improve the measurement statistics. \\
The tomographically reconstructed PMD $\mathcal{P}(\varepsilon,\vartheta,\phi)$, i.e. the density of the superposition wave packet, measured for a chirp parameter of $\phi^{(2)}=\SI{40}{\femto\second^2}$, is shown in Fig.~\ref{fig3}(a). 
It has an intricate spiral shape, suggesting the superposition of multiple symmetry components. Quantitatively, we extract the contributions of these components by 3D Fourier analysis of the PMD in the azimuthal direction. The bar plot in the upper left inset to Fig.~\ref{fig3}(a) shows the resulting amplitude distribution of the symmetry components, determined by integrating the 2D amplitudes $\mathcal{C}_j(\varepsilon,\vartheta)$ in Eq.~\eqref{eq:Fouriersynthesis} over the polar angle and the kinetic energy
\begin{equation}
	\bar{\mathcal{C}}_j=\int\limits_0^\pi\int\limits_0^\infty\mathcal{C}_j(\varepsilon,\vartheta)\varepsilon^2\sin(\vartheta)\mathrm{d}\varepsilon\mathrm{d}\vartheta.
\end{equation}
The amplitude distribution confirms that the PMD consists predominantly of a $c_2$-, a $c_4$- and a $c_6$-symmetric contribution. Odd-numbered symmetry components are not observed due to the symmetrization of the recorded VMI images prior to the tomographic reconstruction. Minor contributions from higher order components are attributed to experimental imperfections, such as noise in the measured VMI images. For visualization purposes, the 3D PMD shown in Fig.~\ref{fig3}(a) has been smoothed by suppression of the higher order symmetry components by a factor of four. By filtering the individual symmetry components and performing the inverse Fourier transform, we obtain the 3D representations shown in Fig.~\ref{fig3}(b)-(d). All three contributions display a spiral-shape in accordance with the SEVs $\mathcal{S}_6$, $\mathcal{S}_4$ and $\mathcal{S}_2$ described in Eqns.~\eqref{eq:SEV6}-\eqref{eq:SEV2}. Fig.~\ref{fig3}(e) shows the azimuthally isotropic DC component $\mathcal{S}_0$.\\
The isosurfaces of the SEVs are color-coded to highlight the phase-jumps observed in the retrieved azimuthal phases $\Phi_j(\varepsilon,\vartheta)$ in the polar direction, similar to the color-coding of the generic SEVs in Fig.~\ref{fig2}. 
To illustrate these phase-jumps, we integrated the 2D phase functions over the photoelectron energy
\begin{equation}
	\bar{\Phi}_j(\vartheta) = \intop_0^\infty \Phi_j(\varepsilon,\vartheta)\varepsilon^2\mathrm{d}\varepsilon
\end{equation}
and plotted the result in the upper right inset to Fig.~\ref{fig3}(a). Phase blanking is applied by fading out the curves to account for decreasing signal amplitudes, i.e. when $\vartheta$ tends to zero. The phase $\bar{\Phi}_6(\vartheta)$ of the $c_6$-symmetric SEV $\mathcal{S}_6$ (red line) is essentially flat. In the polar interval $\vartheta\in[\SI{60}{\degree},\SI{90}{\degree}]$, which contains most of the signal, no phase-jumps are observed. Equation~\eqref{eq:SEV6} shows that the polar part of $\mathcal{S}_6$, given by $P_{3,3}^2[\cos(\vartheta)]$, is strictly positive in this interval. In contrast, the phase $\bar{\Phi}_4(\vartheta)$ of the $c_4$-symmetric SEV $\mathcal{S}_4$ (green line) shows a sharp $\pi$-jump around $\vartheta=\SI{64}{\degree}$, as also observed in previous studies with TL CRCP pulse sequences \cite{Pengel:2017:PRA:043426,Koehnke:2022:JPBAMOP:184003}. The observed phase-jump angle agrees with the sign-change at $\vartheta=\SI{63.4}{\degree}$ of the associated Legendre polynomial $P_{3,1}[\cos(\vartheta)]$ in Eq.~\eqref{eq:SEV4}, which governs the polar part of $\mathcal{S}_4$. The measured phase $\bar{\Phi}_2(\vartheta)$ of the $c_2$-symmetric SEV $\mathcal{S}_2$ (blue line) shows two blurred phase-jumps centered around $\vartheta=\SI{64}{\degree}$ and $\vartheta=\SI{39}{\degree}$, respectively. These phase-jumps result from the superposition of $\Leg{3}{1}^2$ and $\Leg{3}{3}\Leg{3}{1}$ which determine the polar part of $\mathcal{S}_2$, as described by Eq.~\eqref{eq:SEV2}. \\
\begin{figure}[t]
	\includegraphics[width=\linewidth]{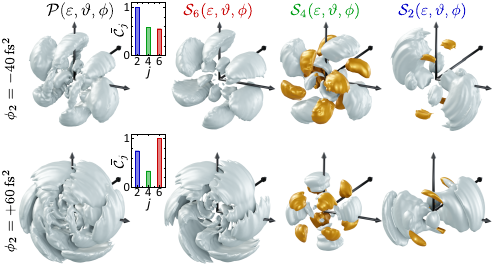}
	\caption{Control of the SEVs via the chirp. The left column shows the reconstructed densities $\mathcal{P}(\varepsilon,\vartheta,\phi)$ of the superposition wave packets measured for $\phi^{(2)}=\SI{-40}{\femto\second^2}$ (top row) and $\phi^{(2)}=+\SI{60}{\femto\second^2}$ (bottom row). The insets show the corresponding amplitude distributions $\bar{\mathcal{C}}_j$ of the SEVs $\mathcal{S}_j$. The SEVs are shown in the second to fourth column. The comparison to the results in Fig.~\ref{fig3} reveals that $(i)$ changing the sign of the chirp reverses the overall sense of rotation of the SEVs and $(ii)$ increasing the magnitude of the chirp tightens the winding of the SEVs. In addition, for increasing chirp, the amplitude of the $c_6$-symmetric SEV is enhanced relative to the amplitudes of the $c_4$- and $c_2$-symmetric SEVs (see main text for details).
	\label{fig4}}
\end{figure}
Next, we control the SEVs by varying the chirp parameter. Figure \ref{fig4} shows the measured and reconstructed PMDs together with the extracted SEVs for the chirp parameters of $\phi^{(2)}=\SI{-40}{\femto\second^2}$ (top row) and $\phi^{(2)}=\SI{60}{\femto\second^2}$ (bottom row). Comparing the top row of Fig.~\ref{fig4} with the results presented in Fig.~\ref{fig3}, we recognize that changing the sign of the chirp from $\phi^{(2)}=+\SI{40}{\femto\second^2}$ to $-\SI{40}{\femto\second^2}$ reverses the rotational sense of the entire SEVs. This property can already be observed in the superposition wave packet, but it is seen most clearly in the SEV $\mathcal{S}_6$. In contrast, increasing the absolute value of the chirp parameter to $\phi^{(2)}=+\SI{60}{\femto\second^2}$ tightens the winding of the SEVs, as seen by comparing the results in the bottom row to Fig.~\ref{fig3}. In addition, the amplitude distribution $\bar{\mathcal{C}}_j$ of the SEVs changes as the chirp is varied, as shown in the insets of Fig.~\ref{fig4}. This observation is rationalized by considering that an increased chirp rate results in an increase in the temporal separation between the spectral components of the same frequency but opposite handedness. As a consequence, the ionization probability of MPI pathways involving both polarization components decreases relative to the pathways involving only a single polarization component. Hence the observed enhancement of the amplitude $\bar{\mathcal{C}}_6$ relative to the amplitudes $\bar{\mathcal{C}}_4$ and $\bar{\mathcal{C}}_2$ shown in the bottom row of Fig.~\ref{fig4}.\\
\begin{figure}[t]
	\includegraphics[width=\linewidth]{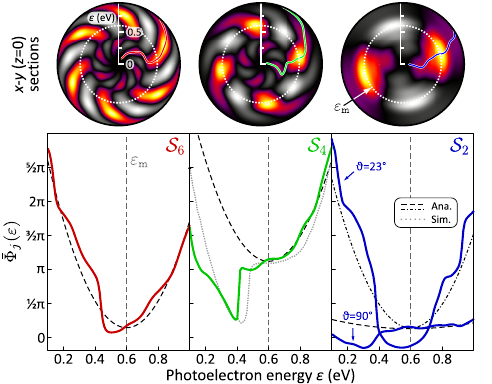}
	\caption{Analysis of the energy-dependent azimuthal phases of the SEVs, measured for a chirp of $\phi^{(2)}=\SI{40}{fs^2}$. The top row shows $x$-$y$-sections ($\vartheta=\SI{90}{\degree}$) through the SEVs. The colored lines are the energy-dependent rotation angles of the spiral arms, given by $\phi_j(\varepsilon)=\bar{\Phi}_j(\varepsilon)/j$. The bottom row shows the retrieved azimuthal phases $\bar{\Phi}_j(\varepsilon)$ of the SEVs $\mathcal{S}_6$ (left), $\mathcal{S}_4$ (middle) and $\mathcal{S}_2$ (right) as colored solid lines. The experimental results are compared to the the analytical model (black dashed lines) and the numerical simulation (gray dotted line). In  all three cases, we observe overall parabolic spectral phase functions, with a blurred phase-jump around $\varepsilon=\SI{0.4}{eV}$. The latter is attributed to the influence of the resonant intermediate state $4p$ in the (1+2) REMPI scenario. \label{fig5}}
\end{figure}
Finally, we discuss the amplitudes and phases of the SEVs measured for $\phi^{(2)}=+\SI{40}{fs^2}$ in more detail. The top row of Fig.~\ref{fig5} shows 2D sections through the $x$-$y$-plane of the SEVs corresponding to the laser polarization plane. 
In this representation, we see qualitatively that the spiral arms start near the center with a clockwise sense of rotation, reverse their rotational sense at the energy $\varepsilon_m$ (indicated with the white dashed circle) and continue outwards with a counter-clockwise sense of rotation.
The geometry of the spiral arms is quantitatively described by the energy-dependence of the retrieved azimuthal phases $\Phi_j(\varepsilon,\vartheta)$.
To obtain a clear picture of the energy-dependence of the azimuthal phases, but avoid averaging over the polar phase-jumps, we integrated the 2D phase functions over a small polar angle-segment $\delta\vartheta\lesssim\SI{10}{\degree}$ centered around the polarization plane ($\vartheta=\SI{90}{\degree}$):
\begin{equation}
	\bar{\Phi}_j(\varepsilon)=\intop_{\SI{90}{\degree}-\delta\vartheta}^{\SI{90}{\degree}+\delta\vartheta}\Phi_j(\varepsilon,\vartheta)\sin(\vartheta)\mathrm{d}\vartheta.
\end{equation}
The ratio $\bar{\Phi}_j(\varepsilon)/j$ of the spectral phase functions and the corresponding symmetry describes the energy-dependent azimuthal rotation angle $\phi_j(\varepsilon)$ of the spiral arms \cite{Kerbstadt:2019:NC:658}. The resulting curves are shown as colored lines in the 2D sections in the top row of Fig.~\ref{fig5}, accurately tracing the respective spiral arms. The phase functions $\bar{\Phi}_j(\varepsilon)$ are plotted in the bottom row (bold solid lines) together with the analytical results predicted by Eqns.~\eqref{eq:phi6_chirp}-\eqref{eq:phi22_chirp} (black dashed lines). Overall, we observe a parabolic energy-dependence of the spectral phases.
In particular, the phase $\bar{\Phi}_6(\varepsilon)$ of the $c_6$-symmetric SEV $\mathcal{S}_6$, shown as red line in the left frame, agrees very well with the analytical model. Around $\varepsilon=\SI{0.4}{\electronvolt}$, the curve falls off more steeply than the model, reminiscent of a blurred phase-jump. Similar features are observed in all three phase functions around this energy. The most pronounced phase-jump is observed in the spectral phase function $\bar{\Phi}_4(\varepsilon)$ of the  $c_4$-symmetric SEV $\mathcal{S}_4$, shown as the green curve in the middle panel of Fig.~\ref{fig5}. To analyze the origin of this phase-jump, we employed a numerical simulation of (1+2)~REMPI with polarization-shaped pulses. The simulation model is based on solving the time-dependent Schrödinger equation for a three-level atom, consisting of the ground state $|4s,m=0\rangle$ and the two resonant states $|4p,\pm1\rangle$, and treating the two-photon ionization from the $4p$-states into $p$- and $f$-type continua by second order time-dependent perturbation theory (for details see \cite{Wollenhaupt:2009:APB:245,Koehnke:2023:NJP:123025}). We used a value of $\SI{0.8}{\radian}$ for the relative phase between the two continua. The numerical result is plotted as a gray dotted curve in the middle frame of Fig.~\ref{fig5}. The phase-jump of $\bar{\Phi}_4(\varepsilon)$ at $\varepsilon=\SI{0.4}{\electronvolt}$ is qualitatively reproduced by the simulation. This result suggests that the phase-jump is due to the influence of the resonant $4p$-state, which is not considered in our analytical model. Further fine-tuning of the spectral phases can be achieved by including higher-lying intermediate states, such as the $3d$-state and the $5s$-state, in the model as well. \\
The retrieved phase of the $c_2$-symmetric SEV $\mathcal{S}_2$,  $\bar{\Phi}_2(\varepsilon)$, is determined by two different phase functions $\varphi_{21}(\varepsilon)$ and $\varphi_{22}(\varepsilon)$, as described by Eq.~\eqref{eq:SEV2}. The analytical phase functions given by Eqns.~\eqref{eq:phi21_chirp} and \eqref{eq:phi22_chirp} are plotted as dashed and dashed-dotted lines, respectively, in the right frame of Fig.~\ref{fig5}. The exact separation of the two phase functions from the retrieved phase is difficult. In our approach, we make use of the polar-dependence of $\Phi_2(\varepsilon,\vartheta)$, exploiting the fact that the amplitudes of the two cosine-terms entering the SEV $\mathcal{S}_2$ vary significantly with $\vartheta$. Inspection of the amplitudes shows that the retrieved phase $\Phi_2(\varepsilon,\vartheta)$ at $\vartheta=\SI{90}{\degree}$ is essentially determined by $\varphi_{21}(\varepsilon)$, whereas at $\vartheta\approx\SI{23}{\degree}$ it is essentially determined by $\varphi_{22}(\varepsilon)$. The corresponding retrieved phase functions $\Phi_2(\varepsilon,\SI{90}{\degree})$ and $\Phi_2(\varepsilon,\SI{23}{\degree})$ are plotted as blue lines in the right frame and agree reasonably well with the analytical results.
The polar-resolved analysis of the retrieved phase $\Phi_2(\varepsilon,\vartheta)$ allows us to approximately disentangle the two phase functions $\varphi_{21}(\varepsilon)$ and $\varphi_{22}(\varepsilon)$ encoded in the SEV $\mathcal{S}_2$.\\

\section{Summary and conclusion\label{Conclusion}}
In this paper, we have reported on the first experimental demonstration of shaped free electron vortices (SEVs) with different rotational symmetries. Initially proposed in \cite{Strandquist:2022:PRA:043110} for atomic single-photon ionization by attosecond pulses, here we used oppositely chirped CRCP fs WLS pulses to drive (1+2) REMPI of potassium atoms. We observed an intricate spiral-shaped 3D PMD consisting of a six-armed ($c_6$-symmetric) SEV superimposed by a four-armed ($c_4$) and a two-armed ($c_2$) SEV. 
Each SEV was characterized by an interference pattern in the shape of a multi-armed spiral whose rotational sense reverses at a distinct photoelectron energy. We controlled the SEVs by variation of the chirp parameter and showed that inverting the sign of the chirp reverses the overall rotational sense of the spirals, while increasing the absolute value of the chirp tightens up the spiral arms. 
Experimentally, we combined $4f$-polarization-shaping of WLS with the VMI-based detection of photoelectron wave packets. The 3D PMD was reconstructed employing photoelectron tomography. 
To disentangle the different SEVs superimposed in the PMD, we exploited their rotational symmetry properties determined by their angular momentum states. 
By Fourier analysis of the PMD in azimuthal direction, we obtained the energy-dependent amplitudes and spectral phases of the SEVs. 
The experimental results are in good agreement with a perturbative analytical model developed in this paper. In the analytical model, the SEV phases encode the optical phases of the pathway-resolved multiphoton spectra. For the case of oppositely chirped pulses studied here, we observed quadratic phase functions with different effective chirp parameters in the different SEVs. In addition, we observed distinct phase-jumps which we attribute to the transient electron dynamics in near-resonant intermediate states, such as the $4p$-state. 
Our interpretation was verified by a numerical simulation of the (1+2) REMPI by oppositely chirped CRCP pulses which included all essential bound states as well as the $f$- and $p$-type continua. \\
Our multiphoton scheme for the creation of SEVs by controlling the spectral phases acquired along multiple MPI pathways becomes even more general by using arbitrarily shaped fs-laser pulses. In this way, it provides a powerful and versatile recipe for controlling the amplitudes and spectral phases of the created angular momentum partial wave packets, allowing to efficiently shape the energy and angular distribution of the generated PMD. In this sense, the scheme opens up new perspectives for rational 3D coherent control of quantum dynamics.
Next, we plan to extend the scheme towards the strong-field regime to study new rapid adiabatic passage scenarios \cite{Vitanov:2001:ARPC:763} in REMPI using intense oppositely chirped CRCP pulses. We are currently applying the 3D Fourier analysis method developed in this paper to the investigation of photoemission time-delays in REMPI \cite{Koehnke:2024}.

\begin{acknowledgments}
	We gratefully acknowledge financially support from the collaborative research program ``Dynamics on the Nanoscale'' (DyNano) and the Wissenschaftsraum ``Elektronen-Licht-Kontrolle'' (elLiKo) funded by the the Nieders{\"a}chsische Ministerium f{\"u}r Wissenschaft und Kultur.
\end{acknowledgments}

\appendix
\section{Analytical description}
In the Appendix, we describe the chirped laser fields used in our experiments, and the corresponding pathway-resolved multiphoton spectra. Section~\ref{App:ChirpedPulses} serves to clarify the notation for the chirped pulse components. The relevant pathway-resolved three-photon spectra used in the analytical model for MPI are derived in Sec.~\ref{App:Pathway_resolved} for equally and oppositely chirped pulses. Finally, we discuss the effective bandwidth and the effective chirp parameter of the multiphoton spectrum for the absorption of multiple chirped fields. 

\subsection{Chirped pulses} \label{App:ChirpedPulses}
The temporal field envelope with the amplitude $\mathcal{E}_0$ and the shape function $g(t)$ is defined as 
\begin{equation}
\mathcal{E}(t) = \mathcal{E}_0 \, g(t).
\label{eq:app_temporal_field}
\end{equation}
We consider a Gaussian-shaped pulse envelope with an intensity FWHM of $\Delta t$
\begin{equation}
 g(t) = \frac{1}{2} \, e^{- \ln(4) \left( \frac{t}{\Delta t} \right)^2},
 \label{eq:app_gaussian_envelope}
\end{equation}
and its spectrum obtained by Fourier transform
\begin{equation}
\tilde{\mathcal{E}}(\omega) = \mathcal{E}_0 \, \Delta t \, \sqrt{\frac{\pi}{8 \ln(2)}} \,  e^{- \ln(4) \left( \frac{\omega}{\Delta \omega} \right)^2}
 = \mathcal{E}_0\tilde{G}(\omega),
\label{eq:app_spectrum_fundamental}
\end{equation}
with a the Gaussian spectral shape function $\tilde{G}(\omega)$ and a FWHM of the power spectral density (PSD) of $\Delta \omega = \frac{\ln(16)}{\Delta t}$. Quadratic spectral phase modulation yields the modulated spectra 
\begin{equation}
\tilde{\mathcal{E}}_{d/u}(\omega) = \tilde{\mathcal{E}}(\omega) \,  e^{- i \frac{\phi^{(2)}}{2} \omega^2 }.
\label{eq:app_mod_spectrum_fundamental}
\end{equation}
A positive (negative) $\phi^{(2)}$ gives rise to an  up-chirp (down-chirp).

\subsection{Pathway-resolved multiphoton spectra} \label{App:Pathway_resolved}
Following the ionization scheme presented in Fig.~\ref{fig1}, the temporal multiphoton field for perturbative (non-resonant) three-photon ionization is given by a product of three fields $\mathcal{E}_{\alpha}(t)$, $\mathcal{E}_{\beta}(t)$ and $\mathcal{E}_{\gamma}(t)$, where $\alpha, \beta, \gamma \in \{\mathrm{u,d}\}$, describing the absorption of different combinations of photons from the two chirped polarization components of the polarization-shaped pulse:
\begin{equation}
\mathcal{E}_{\alpha\beta\gamma}(t) = \mathcal{E}_{\alpha}(t)\mathcal{E}_{\beta}(t)\mathcal{E}_{\gamma}(t).
\label{eq:app_mod_multi_temporal}
\end{equation}
In the experiment, we use chirp parameters of the same magnitude $|\phi^{(2)}|$ but opposite sign for the two components. The more general MPI scenario where each field is chirped arbitrarily is discussed below in Sec.~\ref{app:arbi_chirp}. The corresponding pathway-resolved multiphoton spectra are obtained by Fourier transformation
\begin{equation}
\tilde{\mathcal{E}}_{\alpha\beta\gamma}(\omega) 
= \mathcal{F}  \left[ \mathcal{E}_{\alpha\beta\gamma}(t) \right] 
= \mathcal{E}_0^3 \tilde{G}_{\alpha\beta\gamma}(\omega)e^{- i \frac{\phi_{\alpha\beta\gamma}}{2} \omega^2}.
\label{eq:app_mod_multi_spectral}
\end{equation}
Because the multiphoton spectra have Gaussian shape, we decompose it into the multiphoton spectral shape function $\tilde{G}_{\alpha\beta\gamma}(\omega)$ and the effective multiphoton chirp parameter $\phi_{\alpha\beta\gamma}$. The spectral width $\Delta \omega_{\alpha\beta\gamma}$ of the multiphoton spectral shape function $\tilde{G}_{\alpha\beta\gamma}(\omega)$ is defined analogously to the FWHM of the PSD in the fundamental spectrum in Eq.~\eqref{eq:app_spectrum_fundamental}. In the following we discuss properties of the multiphoton spectra $\tilde{\mathcal{E}}_{\alpha\beta\gamma}(\omega)$  in terms of their spectral widths $\Delta\omega_{\alpha\beta\gamma}$ and effective chirp parameters $\phi_{\alpha\beta\gamma}$. To describe the experiment, it is sufficient to discuss $\tilde{\mathcal{E}}_\mathrm{uuu}(\omega)$  and $\tilde{\mathcal{E}}_\mathrm{uud}(\omega)$ since the other pathway-resolved multiphoton spectra, such as $\tilde{\mathcal{E}}_\mathrm{udu}(\omega)$ or $\tilde{\mathcal{E}}_\mathrm{ddu}(\omega)$ either arise from the change of sign of the chirp parameter or are identical through permutation of $d$ and $u$ in Eq.~\eqref{eq:app_mod_multi_temporal}.

\subsubsection{Equally chirped pulses}\label{app:equal_chirp}
Introducing the dimensionless parameter
\begin{equation}\label{eq:eta}
	\eta= \left[ \ln(16) \, \frac{\phi^{(2)}}{\Delta t^2} \right]^2 =\left[\frac{\Delta\omega^2\phi^{(2)}}{\ln(16)}\right]^2,
\end{equation}
where $\eta \rightarrow 0$ as $\phi^{(2)} \rightarrow 0$, the pathway-resolved multiphoton spectrum for the combination of three up-chirp photons is given by
\begin{equation}
	\tilde{\mathcal{E}}_\mathrm{uuu}(\omega) = \mathcal{E}_0^3\tilde{G}_\mathrm{uuu}(\omega)
	\, e^{- i \frac{\phi_\mathrm{uuu}}{2} \omega^2}
\end{equation}
with the spectral shape function
\begin{align}
	\tilde{G}_\mathrm{uuu}(\omega) = \frac{1}{8}\sqrt{\frac{\pi}{6 \ln(2) } }
	\frac{\Delta t}{(1+i\sqrt{\eta})}e^{- \ln(4) \left( \frac{\omega}{\Delta \omega_\mathrm{uuu}}\right)^2 }
\end{align}
and the effective multiphoton bandwidth and multiphoton chirp parameter
\begin{equation}
\Delta \omega_\mathrm{uuu} = \sqrt{3} \, \Delta \omega 
\quad \mathrm{and} \quad 
\phi_\mathrm{uuu} = \frac{\phi^{(2)}}{3},
\label{eff_para_uuu}
\end{equation}
respectively. For equally chirped pulses the effective parameters are obtained by scaling the parameters of the fundamental pulse by constant factors.
\begin{figure*}[t]
	\includegraphics{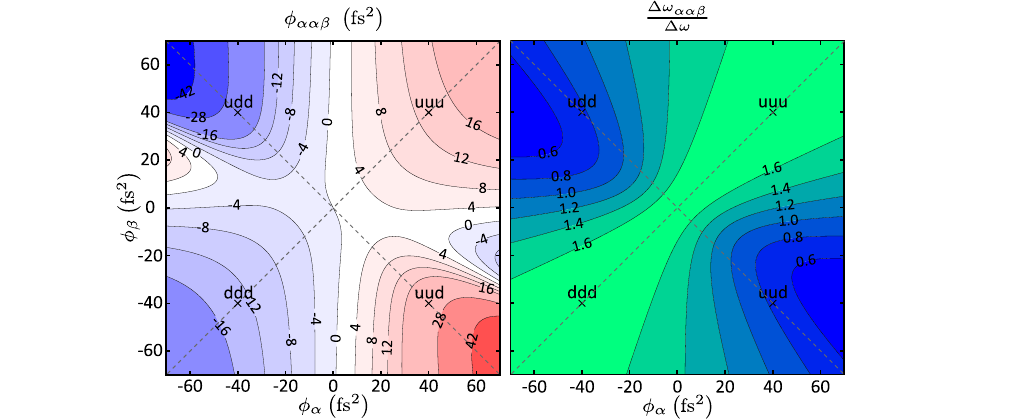}
	\caption{Pathway-resolved multiphoton chirp parameter $\phi_{\alpha\alpha\beta}$ and relative effective multiphoton bandwidth $\frac{\Delta\omega_{\alpha\alpha\beta}}{\Delta\omega}$ of a multiphoton field $\mathcal{E}_\alpha(t)^2\mathcal{E}_\beta(t)$ using a pulse duration of $\Delta t=\SI{6}{\femto\second}$ and the respective chirp parameters $\phi_{\alpha}$ and $\phi_{\beta}$. Each quadrant reflects a different combination of photons from up- and down-chirped fields, i.e., uuu, uud, ddd, udd. In the experiment, chirped fields with the same absolute value $|\phi_{\alpha}|=|\phi_\beta|$ are considered, described by the diagonal lines shown as dashed lines. The effective parameters for the prominent case of $\phi_{\alpha/\beta}=\pm\SI{40}{\femto\second^2}$ in the experimental results are marked by crosses.\label{fig6}}
\end{figure*}
\subsubsection{Oppositely chirped pulses}\label{app:oppo_chirp}
The pathway-resolved multiphoton spectrum for the absorption of two photons from the up-chirped field and one photon from the down-chirp field reads
\begin{equation}
	\tilde{\mathcal{E}}_\mathrm{uud}(\omega) = \mathcal{E}_0^3\tilde{G}_\mathrm{uud}(\omega)
	\, e^{- i \frac{\phi_\mathrm{uud}}{2} \omega^2}.
\end{equation}
The corresponding spectral shape function is given by
\begin{align}
	\tilde{G}_\mathrm{uud}(\omega) =\frac{1}{8}\sqrt{\frac{\pi}{\ln(4)}}\frac{\Delta t}{\sqrt{3+\eta+2i\sqrt{\eta}}}e^{- \ln(4) \left( \frac{\omega}{\Delta \omega_\mathrm{uud}}\right)^2 },
\end{align}
with the effective multiphoton bandwidth and multiphoton chirp parameter 
\begin{equation}
\Delta \omega_\mathrm{uud} = \sqrt{\frac{1}{3}\frac{\eta+9}{\eta+1}} \, \Delta\omega 
\quad \mathrm{and} \quad 
\phi_\mathrm{uud} = \frac{\eta+1}{\eta+9} \, \phi^{(2)},
\end{equation}
respectively. Recalling that $\eta$ defined in Eq.~\eqref{eq:eta} depends on $\Delta\omega$ and $\phi^{(2)}$, the effective parameters describing the multiphoton field for the non-linear absorption of photons from oppositely chirped pulses 
exhibits a non-trivial dependence on both, the effective bandwidth and the multiphoton chirp.
\subsubsection{Arbitrarily chirped pulses}\label{app:arbi_chirp}
So far, we have considered MPI with chirped pulses having an identical magnitude $|\phi^{(2)}|$ of the chirp parameter. However, a more general MPI scenario involves the absorption of multiple ($N$) chirped fields $\mathcal{E}_{k}(t)$, each of which has a different chirp $\phi_k$. By defining
\begin{equation}
	\frac{1}{\Gamma} =  \sum_{k=1}^N  \left[ i \, \phi_k + \frac{\Delta t^2}{\ln(16)} \right]^{-1},
\end{equation}
we obtain the resulting effective multiphoton chirp parameter and the effective bandwidth as
\begin{equation}
	\phi_\mathrm{eff} = {\rm Im} (\Gamma) \quad \mathrm{and} \quad 
	\Delta\omega_\mathrm{eff} = \sqrt{ \frac{ \ln(16) }{ {\rm Re}(\Gamma)   }   },
\end{equation}
respectively. As an example, we consider three-photon ionization with the arbitrarily chirped fields $\mathcal{E}_{\alpha}(t)$, $\mathcal{E}_{\beta}(t)$ and $\mathcal{E}_{\gamma}(t)$. Here, the effective chirp parameter $\phi_{\alpha\beta\gamma}$ and the effective bandwidth $\Delta\omega_{\alpha\beta\gamma}$ are functions of all three chirp parameters and of the TL pulse duration $\Delta t$.
We illustrate the effect of the mixing of two pulses with different chirps in the three-photon field $\mathcal{E}_\alpha^2(t)\mathcal{E}_\beta(t)$ by plotting the effective multiphoton chirp parameter $\phi_{\alpha\alpha\beta}$ and relative effective bandwidth $\frac{\Delta\omega_{\alpha\alpha\beta}}{\Delta\omega}$ as a function of the two chirp parameters $\phi_\alpha$ and $\phi_\beta$ in Fig.~\ref{fig6}.
As expected from Eq.~\eqref{eff_para_uuu}, the effective chirp parameter on the diagonal, where $\phi_\alpha=\phi_\beta=\phi^{(2)}$, is $\phi^{(2)}/3$. Non-intuitively, the effective multiphoton chirp parameter $\phi_{\alpha\alpha\beta}$ of the multiphoton field $\mathcal{E}_\alpha^2(t)\mathcal{E}_\beta(t)$ is positive, although the field $\mathcal{E}_{\alpha}(t)$, contributing two photons, has a large negative chip parameter $\phi_\alpha$, whereas the field  $\mathcal{E}_{\beta}(t)$, contributing only one photon, has only a small positive chip parameter $\phi_\beta$. 

\bibliography{ultra_db.bib}

\end{document}